# EPE

## *The Extreme Physics Explorer*

A non-proprietary Mission Concept available for presentation to NASA, providing
High Area, High Resolution Imaging Spectroscopy and Timing with Arcmin Angular Resolution

Submitted in response to NASA 2011 RFI NNH11ZDA018L
'Concepts for the Next NASA X-ray Astronomy Mission'

Michael Garcia, Astrophysicist
Smithsonian Astrophysical Observatory
1-617-495-7169
garcia@cfa.harvard.edu

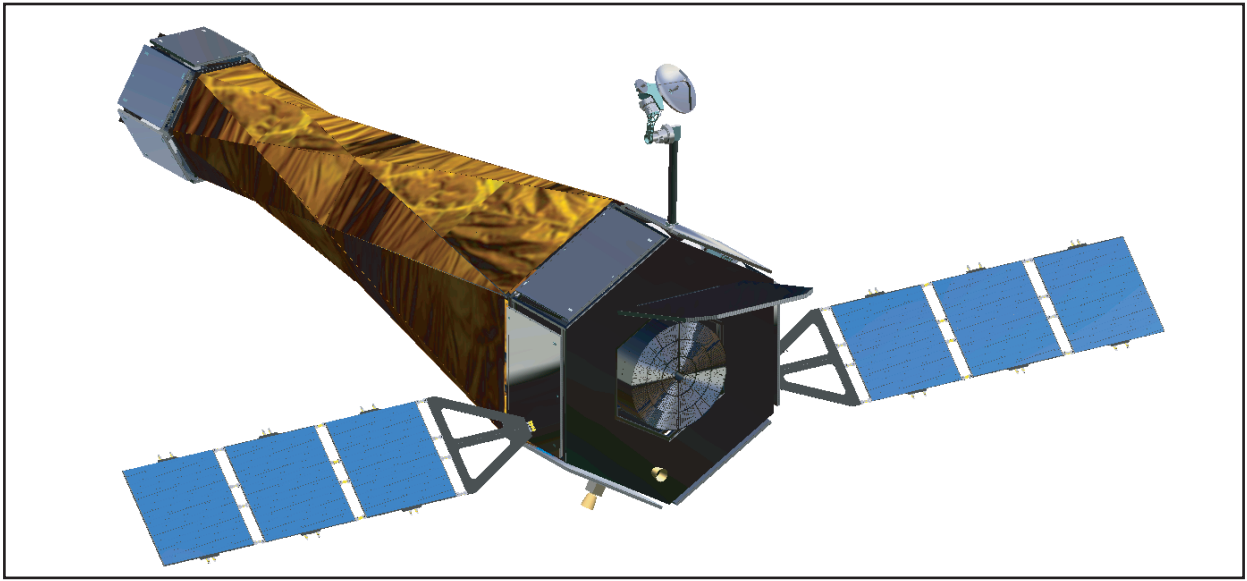

Michael Garcia, Martin Elvis, Jay Bookbinder, Laura Brenneman, Esra Bulbul, Paul Nulsen,
Dan Patnaude, Randall Smith,  (SAO)
Simon Bandler, Takashi Okajima, Andy Ptak (NASA/GSFC)
Enectali Figueroa-Feliciano, Deepto Chakrabarty (MIT)
Rolf Danner, Dean Daily (Northrop Grumman)
George Fraser, Richard Willingale (University of Leicester)
Jon Miller (Univesity of  Michigan), T.J. Turner (UMBC)
Guido Risalti (Arcetri Observatory),  Massimiliano Galeazzi (University of Miami)

# 1. SUMMARY

The Extreme Physics Explorer (EPE) is a mission concept that will address fundamental and timely questions in astrophysics which are primary science objectives of IXO. The reach of EPE to the areas outlined in NASA RFI NNH11ZDA018L is shown in Table 1. The dark green indicates areas in which EPE can do the basic IXO science, and the light green areas where EPE can contribute but will not reach the full IXO capability.

To address these science questions, EPE will trace orbits close to the event horizon of black holes, measure black hole spin in active galactic nuclei (AGN), use spectroscopy to characterize outflows and the environment of AGN, map bulk motions and turbulence in galaxy clusters, and observe the process of cosmic feedback where black holes inject energy on galactic and intergalactic scales.

EPE gives up the high resolution imaging of IXO in return for lightweight, high TRL foil mirrors which will provide >20 times the effective area of ASTRO-H and similar spatial resolution, with a beam sufficient to study point sources and nearby galaxies and clusters. Advances in micro-calorimeters allow improved performance at high rates with twice the energy resolution of ASTRO-H. A lower TRL option would provide 200 times the area of ASTRO-H using a micro-channel plate optic (MCPO) and a deployable optical bench. Both options are in the middle range of RFI missions at between $600M and $1000M.

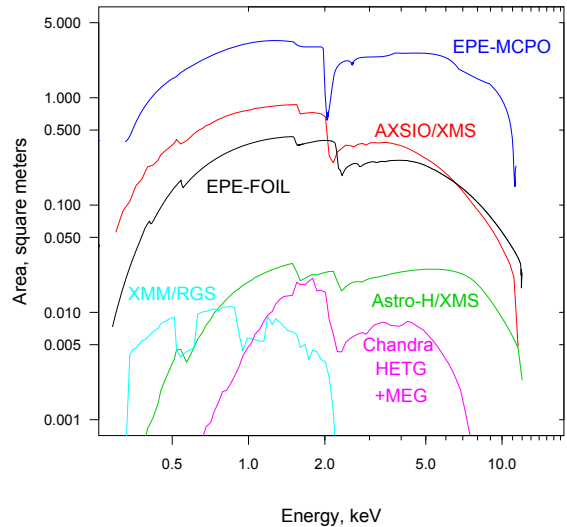

Figure 1. The EPE effective area will be >20x ASTRO-H and only slightly below AXSIO, allowing detailed studies over a wide range of IXO science.

The EPE foil optic has direct heritage to ASTRO-H, allowing robust cost estimates. The spacecraft is entirely off the shelf and introduces no difficult requirements. The mission could be started and launched in this decade to an L2 orbit, with a three-year lifetime and consumables for 5 years. While ASTRO-H will give us the first taste of high-resolution, non-dispersive X-ray spectroscopy, it will be limited to small numbers of objects in many categories. EPE will give us the first statistically significant samples in each of these categories.

| RFI Table 1: Primary IXO Science Objectives Addressed by Extreme Physics Explorer | | |
|---|---|---|
| **Science Question** | **IXO Measurement** | **EPE Reach** |
| What happens close to a black hole? | Time resolved high resolution spectroscopy of stellar mass and ~20 supermassive black holes | Time resolved high resolution spectroscopy of stellar mass and >dozen (foil) or ~5 dozen (mcpo) supermassive black holes |
| When and how did SMBH grow? | Measure the spin in 300 supermassive black holes within z < 0.2 | Measure the spin in >40(foil) to ~1000(mcpo) supermassive black holes |
| How does large scale structure evolve? | (i) Find the missing baryons via WHIM absorption line spectroscopy using AGN as illumination sources. | Detection threshold ~70% of IXO; expect detections on 20 lines of sight and of ~100 absorbers. |
| | (ii.) Measure the mass and composition of ~500 clusters of galaxies at redshift < 2 | Measure the mass and composition of the 100 brightest nearby clusters from REFLEX catalog. |
| Connection between SMBH and large scale structure ? | Measure the metallicity and velocity structure of hot gas in galaxies & clusters. | Spatially resolve and measure abundances and wind velocities in several dozen starburst galaxies; measure velocity and turbulence in ~6 nearby cluster bubbles. |
| How does matter behave at very high density? | Measure the equation of state of neutron stars through (i.) spectroscopy and | Unique timing/spectral capability allows phase-binning on ms period rotation rates. |
| | (ii.) timing | 80% throughput at several Crab fluxes – no diffusing optic, so area matches IXO above ~5 keV. |

## 2. SCIENCE

The extragalactic X-ray sky is dominated by two kinds of sources: accreting supermassive black holes (SMBH) in galactic nuclei, comparable in size to the Solar System, and clusters of galaxies, more than a million light years across. The energy liberated by growing black holes influences the infall of gas in galaxies and clusters, while some analogous process, still poorly understood, ties the growth of black hole mass to a fixed fraction of its host galaxy's bulge.[1,2,3,4] The remarkable link between SMBH, galactic nuclei, and the largest gravitationally bound structures in the universe (clusters) implies that a two-way connection, called "feedback", is a key ingredient for understanding them all.

The driving science goals of EPE are to measure the energetics and dynamics of the hot gas in large cosmic structures and to determine the properties of the extreme environment and evolution of black holes, in order to understand the connection between SMBH, galaxies, and large scale structure. EPE will also constrain the equation of state of neutron stars and track the dynamical and compositional evolution of interstellar and intergalactic matter in the local universe.

### *What Happens Close to a Black Hole?*

Black holes harbor the strongest gravitational fields and are among the most extreme environments in the Universe. EPE's capabilities will allow us to answer the questions: What are the effects of strong gravity close to a black hole event horizon? How do black holes grow, evolve, and influence galaxy formation?

The observational consequences of strong gravity can be seen close to the event horizon, where the extreme effects of General Relativity (GR) are evident in the form of gravitational redshift, light bending, and frame dragging. The spectral signatures needed to determine the physics of the accretion flow into the black hole are only found in X-rays. EPE will allow us to observe orbiting features from the innermost accretion disk where strong gravity effects dominate (Astro2010 White Paper (APW): Spin and Relativistic Phenomena around Black Holes, Brenneman et al.). Observations of the brightest and most massive SMBH with XMM-Newton have revealed evidence of "hot spots" on the disk that light up in the Fe-Kα line, allowing us to infer their motions.[6] Each parcel of gas follows a nearly circular orbit around a black hole. We note that nearly all Seyfert~1 SMBH show Fe-Kα lines[6,32] but tracing hot spot motions on sub-orbital timescales requires large effective area over the 3 keV to 7 keV band which contains the GR broadened and shifted Fe-Kα line. The area must be sufficient to detect the Fe-Kα line above the continuum and determine its velocity centroid within a fraction of an orbital timescale. The emission from these hot spots appears as "arcs" in the time-energy plane. GR makes specific predictions for the form of these arcs, and the ensemble of arcs reveals the mass and spin of the black hole and the inclination of the accretion disk. Deviations from the GR predictions will create apparent changes in these parameters as a function of time or hot spot radius. EPE will enable the first orbitally time-resolved studies of >1(5) dozen SMBH with the foil (MCPO) optics. The foil optic allows studies over a factor of 20 in SMBH mass, while the MCPO increases this to a factor of 1000.

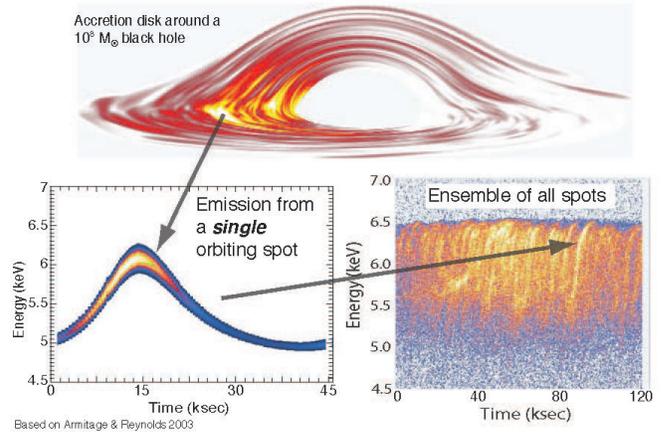

*Figure 2. Simulation of the Fe-Kα lines in a SMBH accretion disk, showing ability of EPE to trace hot spot orbits.[5]*

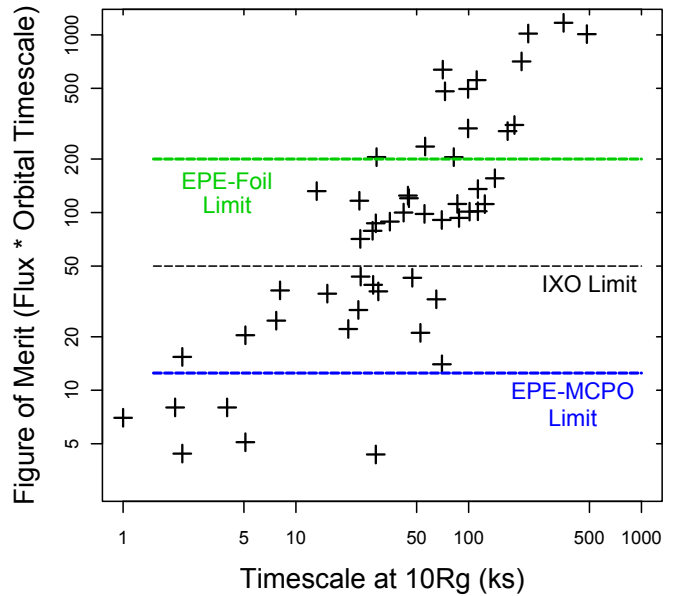

*Figure 3. The orbital timescale vs. figure of merit for measuring motions of hot spots in SMBH accretion disks. The ASTRO-H limit in these units is ~3000, limiting it to the heaviest SMBH, while EPE can reach >dozen. Fluxes are for Seyfert 1s from the BAT 58 month catalog.*

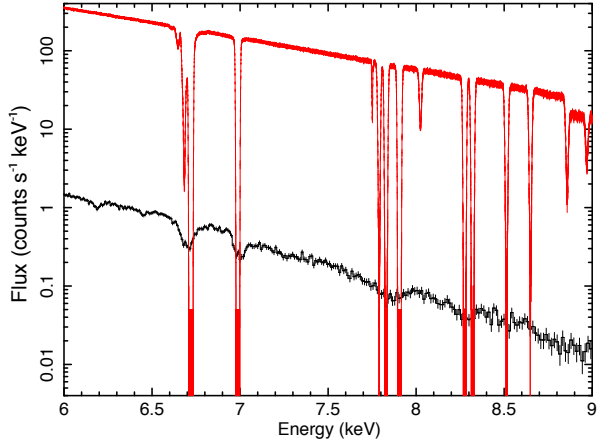

*Figure 4. The Chandra 50ks HETGS spectrum of GRO J1655-40 in black, and a 50ks EPE exposure in red.*

These observations provide a direct probe of the physics of strong gravity. The lower angular resolution of EPE as compared to IXO imposes no limitations on the reach of EPE - the sole requirements are for effective area and spectral resolution as background and source blending are negligable.

**Disk winds from black holes** are likely an important source of hot gas in galaxies with small and moderate bulge components. This is 'feedback' and it can inhibit star formation, and thus influence galaxy evolution. The hottest component of disk winds carries the bulk of the mass flux, and this component is traced through blue-shifted X-ray absorption lines. Observations with Chandra and XMM-Newton have started to probe these winds in both AGN and stellar-mass black holes. A particularly good example is the Chandra spectrum of the black hole binary GRO J1655-40; in this case, the wind is found to originate within 500 Schwarzshild radii of the black hole, and to be driven partly by magnetic pressure from the disk. Currently, this is the best spectrum of a black hole disk wind -- from a black hole of any mass -- in the Fe K band[7]. Whereas scattering in the HETGS makes it difficult to determine if the lines are black, the EPE calorimeter concentrates the line flux, and reveals the absorption in terrific detail (Figure 4). EPE spectra of black holes will be revolutionary, owing to the powerful combination of collecting area and resolution in critical Fe K band (AWP: Stellar-Mass Black Holes and Their Progenitors, Miller et al.).

## When and how did supermassive black holes grow?

SMBHs are a critical component in the formation and evolution of galaxies. Future observatories including JWST, ALMA and 30m-class ground-based telescopes will observe the starlight from galaxies out to the highest redshift. Gas dynamical simulations of the first galaxies predict a period of intense star formation and obscured accretion, driven by a rapid sequence of mergers.[8] The light from this obscured accretion, in particular that onto the central black holes, is most naturally observed in the X-ray band. EPE can constrain these evolutionary models by measuring the black hole spin in four independent ways: relativistic disk line spectroscopy, reverberation mapping, disk hot spot mapping, and power spectral analysis (AWP: Spin and Relativistic Phenomena around Black Holes, Brenneman et al.). In SMBHs, the spin can be changed by either accretion or merger. The current spin distribution is a record of the relative importance of mergers versus accretion in the growth history of black holes. The key observational signature is the Fe-K$\alpha$ emission line, produced via the illumination of the disk by the primary X-ray continuum and distorted in energy and strength by the gravitational field and relativistic motions around

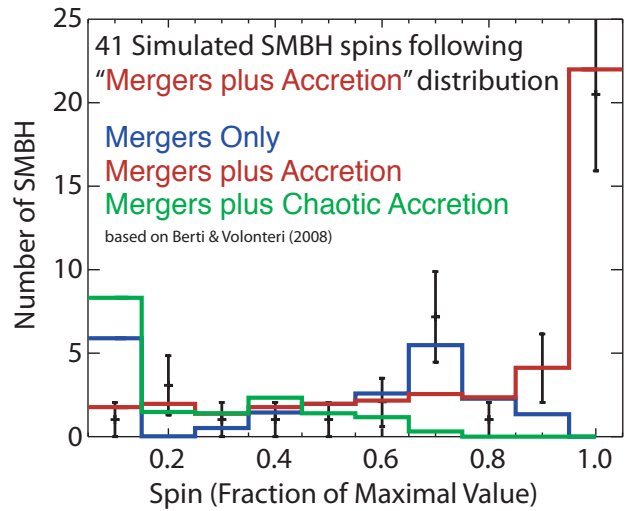

*Figure 5. The data points with error bars are drawn from the red distribution, and show that 41 measurements is sufficient to robustly distinguish between the theoretical distributions. The EPE foil optic configuration will allow at least this many measurements.*

the black hole.

If we simply scale the AGN surface density used for the IXO studies submitted to Astro2010, we estimate that EPE (foil) studies will determine the spin of >40 SMBHs (see Figure 5). This is sufficient to robustly distinguish merger from accretion models and providing a new constraint on galaxy evolution. However, since the Astro2010 IXO studies, the ongoing SWIFT/BAT survey has revealed many new AGN which are bright in the 2-10 keV band crucial for measuring Fe-Kα line shapes. Measuring spin to 1% to 10% requires $10^6$ to $10^5$ integrated counts over 2-10 keV, with the larger number being required for those sources with complex warm-absorbers. The numbers of AGN with complex vs. simple absorption is uncertain, and indeed it will take a large area calorimeter mission like EPE to determine this. However, the least absorbed Seyferts are the Seyfert 1.0s, and there are 54 of these in the BAT-9 month catalog[31] and 198 in the BAT-58 month catalog[32]. Given the BAT survey results it is likely that EPE (foil) will be able to determine spin to better than 10% in >100 AGN given a dedicated 10ms observing program. The 2-10 keV count rate provided by the MCPO is ~8x higher for a typical AGN spectrum, indicating that it would allow spin measurement for a substantial fraction of the 1092 AGN in the BAT-58 month catalog[32].

### *How does large scale structure evolve?*

The extraordinary capabilities of EPE will reveal the major baryonic component of the Universe, in clusters, groups and the intergalactic medium (IGM), and the interplay between these hot baryons and the energetic processes responsible for cosmic feedback. EPE will open a new era in the study of galaxy clusters by directly mapping the gas bulk velocity field and turbulence. EPE's sensitivity will enable us to confront key questions: Where are the Missing Baryons in the Universe? How does Cosmic Feedback Work? How did Large Scale Structure Evolve?

### The Cosmic Web of Baryons

Less than 10% of the baryons in the local Universe lie in galaxies as stars or cold gas, with the remainder predicted to exist as a dilute gaseous filamentary network—the cosmic web. Some of this cosmic web is detected in Lyα and OVI absorption lines, but half remains undetected. Growth of structure simulations predict that these "missing" baryons are shock heated up to temperatures of $10^{6-7}$ K in unvirialized cosmic filaments and chemically enriched by galactic superwinds,[9] so forming the 'warm-hot intergalactic medium' (WHIM).

Despite local success in finding hot gas in the halo of the Milky Way, observations with the grating spectrometers on XMM-Newton and Chandra have not yielded conclusive proof for the existence of the hot cosmic web at z > 0.10 Despite the lower spectral resolution of the EPE calorimeter at 0.6 keV as compared to IXO and AXSIO grating spectrometers, the increased area gives 70% of the detection capability of that grating (AWP: The

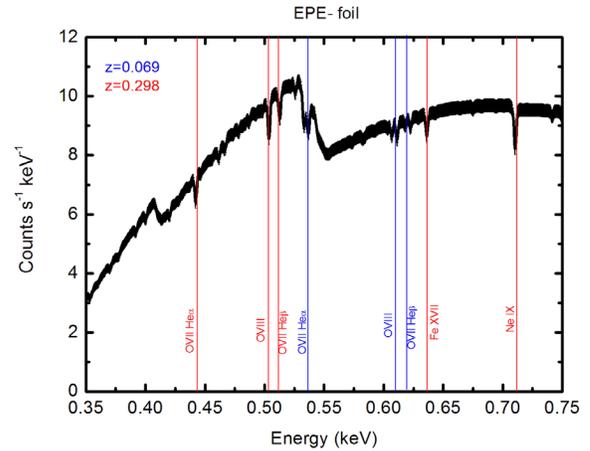

*Figure 6. Typical absorption spectrum in the direction of a bright AGN due to the intervening WHIM gas. Two WHIM filaments are clearly seen at redshifts of 0.069 and 0.298.*

Cosmic Web of Baryons, Bregman et al.). During a three year lifetime of the EPE mission, we expect about 20 observations of 500ks each of bright AGN dedicated to studying the WHIM, plus a few additional 'serendipitous' WHIM detections on long observations of bright AGN designed to look at Fe-Kα line variability. A typical observation with the EPE is shown in Fig. 6 for the baseline foil mirror configuration. The plot shows a reference AGN with galactic absorption and simulated WHIM absorption[11,12] for an assumed exposure time of 500 ks. While bright AGNs are quite common, most of them are nearby, hence probing, relatively short lines of sight. However, Cross-correlating the VERONCAT catalog of quasars and AGNs[13] with the ROSAT All-Sky Survey Bright Catalogs[14] finds 16 objects (mostly BLLac) at z > 0.315 with a 200 ks integrated flux (fluence) in the (0.3–2) keV range above the $10^{-6}$ erg cm$^{-2}$ level represented in Figure 6. This therefore matches our planned program of ~20 observations well. The mean redshift of these objects is ~0.5. The MCPO would allow this program to be carried out in 1ms rather than 10ms, or alternatively along many more lines of sight. Identification of additional suitable AGN would be possible once the eROSITA all-sky survey was complete.

**Mass Loss:** Most galaxies, in fact, have lost more than 2/3 of their baryons, relative to the cosmological ratio of baryons to dark matter.[16] These missing baryons are probably hot, but we do not know if they were expelled as part of a starburst phase galactic wind, or were pre-heated so that they simply never coalesced. X-ray absorption line observations with EPE will, for the first time, identify the location and metallicity of these baryons in the Local Group from the line centroids and equivalent widths of hot C, N, and O ions, revealing a crucial aspect of galaxy formation (Figure 7 and AWP: The Missing Baryons in the Milky Way and Local Group, Bregman et al.). ASTRO-H plans to measure these winds in a few starbursts, but with EPE we can spatially resolve the flows in more than a dozen and can measure the integrated wind properties in another two dozen, yielding statistically meaningful samples (SWP: Starburst Galaxies: Outflows of Metals and Energy into the IGM, and technical supplement, Strickland et al.).

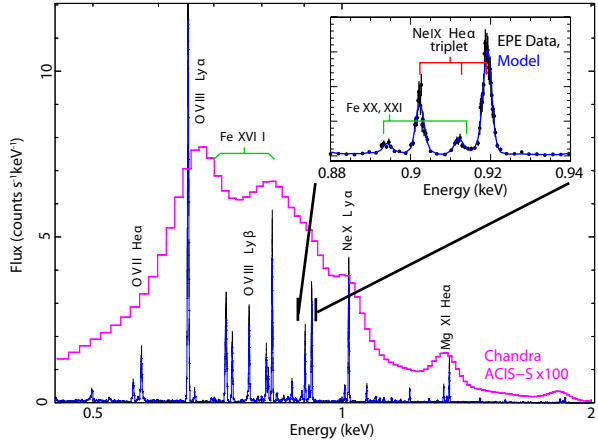

*Figure 7. Simulated EPE spectrum of a 1 arcmin region of a starburst superwind (blue) and the equivalent spectrum at current CCD resolution. EPE resolves the triplets needed to provide velocity, abundance, and temperature diagnostics not available with CCDs.*

**Mass and Composition of Galaxy Clusters**

While EPE will not be able to measure the mass and composition of the 500 to 1000 clusters up to z=2 as IXO could, it will be able to measure these properties in nearby bright clusters, including the >100 brighter than the Bullet Cluster in the REFLEX catalog[17]. This compares to the dozen or so planned for ASTRO-H. These nearby clusters have sizes easily resolved by EPE, as shown Figure 8. The EPE detector background level (foil optics) is reached at ~1/3(r_500), so mass and composition measurements are possible for all of these clusters. The ability of EPE to measure turbulence in the cluster gas will allow the level of non-thermal pressure support to be quantified, therefore removing one of the remaining outstanding issues in cluster mass measurements, and improving the utility of cluster X-ray measurements as cosmological probes (Figure 9).

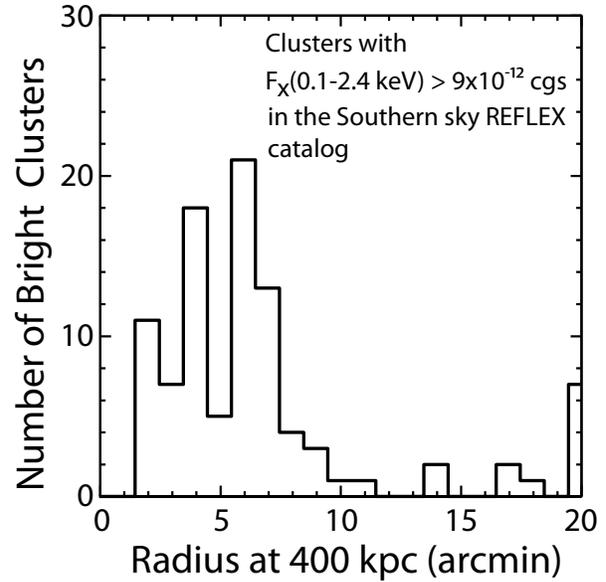

*Figure 8. The radius at ~1/3(r_500) for 100 nearby, bright clusters from the REFLEX catalog. EPE will spatially resolve and determine accurate mass and composition for all of these.*

**Enrichment:** Measuring the metal content and abundance pattern of the IGM with EPE will show when and how the metals are produced, in particular the relative contribution of Type Ia and core-collapse supernovae, and the stellar sources of iron, carbon and nitrogen. Precise abundance profiles from EPE measurements will constrain how the metals produced in the galaxies are ejected and redistributed into the intra-cluster medium.

### What is the connection between SMBH and large scale structure?

**Feedback:** Energetic processes around black holes result in huge radiative and mechanical outputs which can potentially have a profound effect on their larger scale environment in galaxies, clusters and the intergalactic medium (AWP: Cosmic Feedback from Massive Black Holes, Fabian et al.). The black hole can heat surrounding gas via its radiative output, and drive outflows via radiation pressure. Mechanical power emerging in winds or jets can also provide heating and pressure. This is seen in the large bubbles or acoustical ripples blown into the IGM (e.g., MS0735.6+7421[18] and Perseus[19]). EPE will be able to map the velocities and turbulence imparted to the IGM by these processes (Figure 9). For outflows that are radiatively accelerated in AGN, EPE observations will determine the total column density and flow velocity, and hence the kinetic energy flux and the true importance of feedback.

     

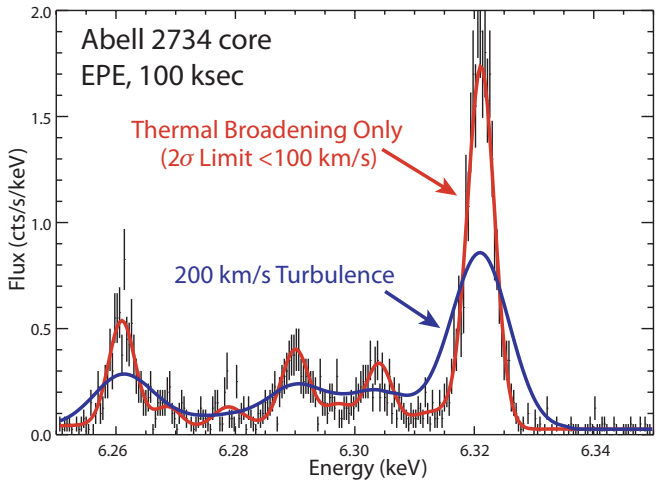

*Figure 9. Simulation of the Fe XXV line obtained from 100 ksec EPE (foil) spectrum of A2734. The thermal broadening of the emission lines alone (red) will be easily distinguished from turbulence of 200 km/s (blue).*

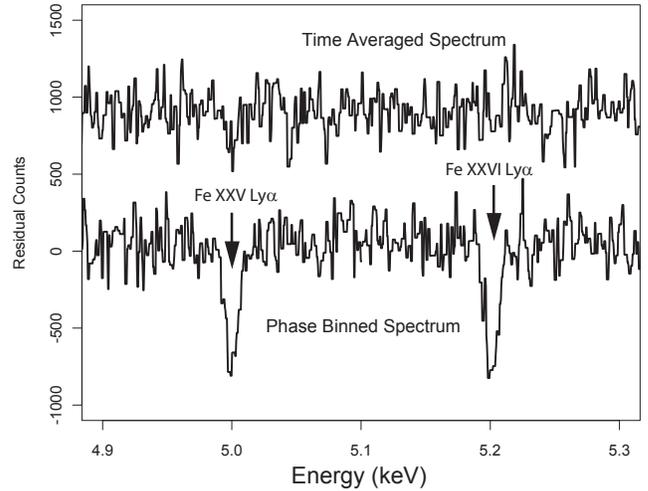

*Figure 10. Simulated 100s (foil) phase-binned spectrum from a typical ~1 Crab burster.*

**Cooling Flows:** In the centers of many galaxy clusters, the radiative cooling time of the X-ray-emitting gas is much shorter than the age of the system. Despite this, the gas there is still hot. Mechanical power from the central AGN acting through jets is thought to compensate for the energy lost across scales of tens to hundreds of kpc. EPE will map the gas velocity across the largest ~half dozen galaxy clusters[20] to an accuracy of ~100 km/s, revealing how the mechanical energy is spread and dissipated.

### How does matter behave at very high density?

Neutron stars (NSs) have the highest known matter densities in nature, utterly beyond the densities produced in terrestrial laboratories. The appearance of exotic excitations and phase transitions to strange matter have been predicted, but these predictions are uncertain due to the complexity of Quantum Chromodynamics (QCD) in this high-density regime. These uncertainties lead to widely differing equations of state, each of which imply a different neutron star radius for a given mass.[21] EPE will determine the mass-radius relationship a ~dozen neutron stars of various masses with four distinct, redundant and complementary methods: (1) the gravitational redshift and (2) Doppler shift and broadening of atmospheric absorption lines, (3) pulse timing distortions due to gravitational lensing, and (4) pressure broadening of line profiles, all enabled by high resolution spectroscopy and energy-resolved fast timing (AWP: The Behavior of Matter Under Extreme Conditions, Paerels et al.).[22,23] The most straightforward method is measuring the gravitational redshift imprinted on narrow atmospheric atomic (ie, He and H-like Fe[23]) lines via binning the spectra at the rotational period of busting NSs, which removes the Doppler broadening from the atmospheric hot spot where the X-ray burst originates. The EPE calorimeter has the unique combination of high rate capacity, large area, sub-ms timing, and superb energy resolution needed for this task. A simulated spectrum from a typical burster (Figure 10) shows the detectability of narrow atmospheric lines in a 100s integration. The top curve results from a simple time integration of the spectrum, and the bottom from phase binning the spectra at the spin period with a (trial) velocity amplitude corresponding to the surface velocity of the rapidly spinning (few ms spin period) neutron star. We note that for slowly spinning NS like Terzan 5 X-2[33] Doppler broadening is small and one can detect the lines in the time averaged spectrum. Shown are the Lyman-alpha lines of H-like and He-like Fe, redshifted by z=0.3. Note that this method alone simultaneously yields the mass and radius. The lines become detectable with integrations greater than 20s. There are a ~dozen bursters suitable for these measurements with EPE. The high count rate capable XMS for EPE gives 80% throughput up to rates of 70,000 c/s in the integrated beam, equivalent to more than 3 Crab. Because the EPE XMS does not need the diffusing optic envisioned for IXO, the effective area above ~5 keV matches IXO. The smaller effective area of ASTRO-H strongly limits its effectiveness for this science topic, as one needs sufficient area to detect the pulsations and therefore determine the phase of the burst hot spot.

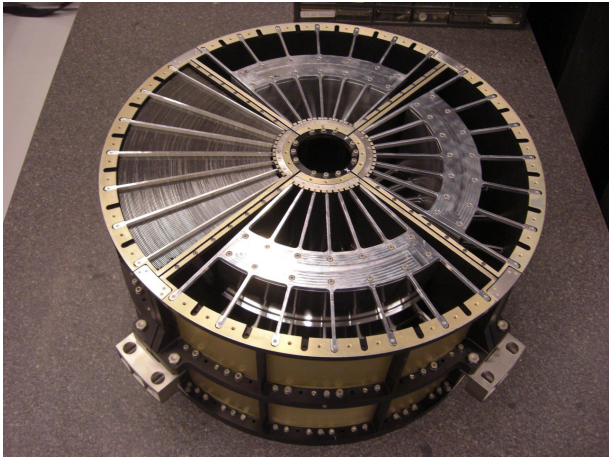

*Figure 11. This Engineering Test Model of the ASTRO-H foil mirror has demonstrated 1.1 arcmin PSF in the one quadrant fully populated.*

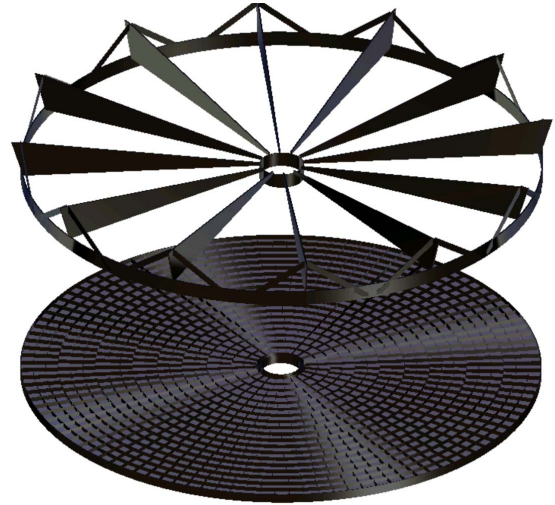

*Figure 12. Possible layout of optional MCPO and support structure.*

## 3. TECHNOLOGY OVERVIEW

### Observatory Overview

The Extreme Physics Explorer is will be placed via direct insertion into an 800,000 km semi-major axis halo orbit around the Sun-Earth L2 libration point using a Falcon 9. The baseline mirror is a larger version of the ASTRO-H foil mirror with a 10m focal length 1 arcmin PSF and fixed optical bench, therefore being a dramatic simplification from IXO and substantially lighter than AXSIO with robust heritage and high TRL (7 to 8). A dramatic increase in effective area could be achieved via a low-TRL (2 or 3) option using a micro-channel plate optic and a TRL-7 extending mast to fit within the launch vehicle faring and give a 40m focal length. The mission design life is three years, with consumables sized for 5 years. The observatory design borrows heavily from IXO and AXSIO and therefore builds on studies performed over the last decade by NASA, ESA, and JAXA, and has strong heritage from previous space flight missions.

### Payload Overview

**The baseline foil mirror** provides ~5,000 cm$^2$ at 1 keV with an ~arcmin level (half power diameter, HPD) angular resolution and light weight. Foil mirror X-ray optics using aluminum substrates, such as the ones on board ASCA, Suzaku, ASTRO-H, have demonstrated excellent area to mass ratios. The Suzaku X-ray telescope achieved 28 cm$^2$/kg, which is 70 times larger than Chandra and 11 times larger than XMM-Newton. The angular resolution of Suzaku is ~2 arcmin (HPD). The ASTRO-H Soft X-ray telescope (SXT) being developed at NASA's Goddard Space Flight Center (GSFC) has the same thin aluminum foil reflectors as Suzaku, but with a larger diameter of 45 cm and a longer focal length of 5.6 m$^{24}$. After many minor refinements in reflector production and positioning in the housing, the SXT Engineering Test Unit recently built at GSFC has archived ~1.1 arcmin angular resolution (HPD), with 15 cm$^2$/kg. The proposed EPE foil mirror builds on these proven successes (the reflectors are TRL-9), to provide the required large effective area with ~arcmin level HPD, low mass and low cost.

ASTRO-H SXT has 563 cm$^2$ at 1 keV with a 0.45 m diameter. For EPE the diameter is extended to 1.3 m, which produces 5,432 cm$^2$ at 1 keV with a reflector mass of 97 kg. Since the diameter is larger than that of ASTRO-H, we estimate the mass of the housing structure to be ~100 kg (~50% of total mass, 197kg). To be conservative we are carrying the 317 kg NGAS mass estimate in the final rack-up. The reflectors will use exactly the same aluminum foils (thickness 150-300 μm) and epoxy replicated gold reflecting surface as Suzaku. The effective area of 5,432 cm$^2$ is based on an axial reflector length of 20 cm. However the Suzaku/ASTRO-H reflector has a 10 cm axial length. If the axial length is doubled it could degrade the angular resolution. In order to keep the same reflector size as Suzaku/ASTRO-H we propose to split each of the primary and the secondary stages axially into two segments, so that the combined mirror has four axial stages, two for the primary stage and the other two the secondary, but still two reflections. The larger radius reflectors beyond the ASTRO-H size (~50 cm diameter) can be divided into eight azimuthal segments instead of four Suzaku/ASTRO-H segments, which also keeps the reflector size similar to Suzaku/ASTRO-H and hence retains the image quality and high TRL. The 1.3 m mirror requires 10,000 reflectors fabricated (including realistic acceptance rate based on Suzaku/ASTRO-H) and we have experience producing >10,000 reflectors for Suzaku. Note that the ASTRO-H mirrors are currently at TLR-8 and will be TRL-9 (flight proven) shortly.

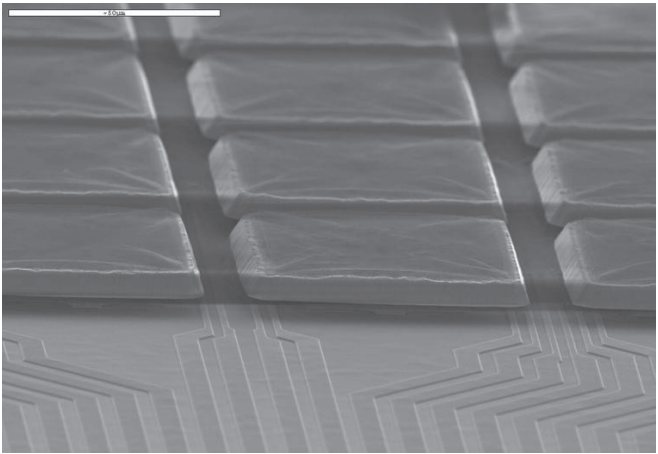
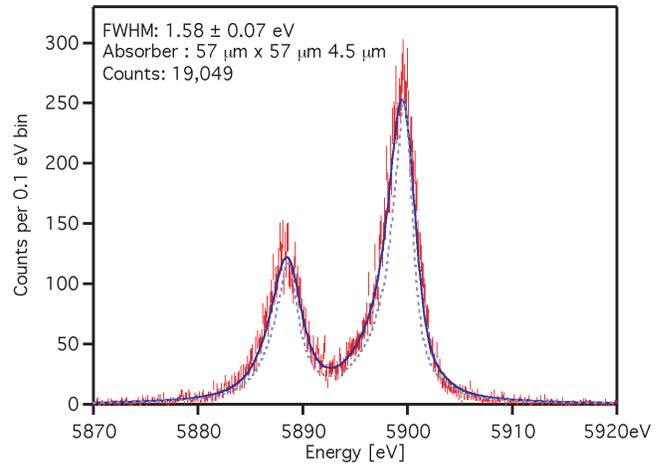

*Figure 13. Left: Scanning electron micrograph image of part of a close-packed 32x32 array of pixels on a 75μm pitch. Right: Performance of a single pixel of this design for 6 keV X-rays. The light blue dashed line is the intrinsic $K\alpha_1$, $K\alpha_2$ line-shape from the Mn source, and the solid dark blue line is the best fit to the data consistent with a Gaussian broadening of 1.58 eV FWHM.*

The only study that is required prior to fabricating the EPE mirror is to establish a way of producing large forming and replication mandrels for the reflector fabrication. The ASTRO-H forming mandrels are conical aluminum cylinders machined by a precision lathe and have <1 μm P-V axial figure. A process to handle the larger EPE forming mandrels on the lathe needs to be established. The Suzaku/ASTRO-H replication mandrels are standard glass cylinder tubes from the German company Schott and we anticipate no problems obtaining the larger cylinders needed for EPE.

**The MCPO enhancement** uses the square pore optics produced by Photonis Inc, in the Wolter-1 geometry used in the Bepi-Columbo MIX-T optics[25]. The mirror could be built up from 1cm square plates, or from ~10 cm wedge shaped tiles, and has a 4.2 m diameter. This second option is shown here, with 19 petal rings required, each consisting of 1344 MCPO. The main remaining technological hurdle with these is decreasing the surface roughness of the reflector walls in order to retain high reflectivity above 5 keV. A promising technique is magnetically assisted finishing[26] but this has not yet been tried on MCPO. Currently this mirror is TRL 2 to 3.

**The X-ray Microcalorimeter Spectrometer (XMS)** provides high spectral resolution, non-dispersive imaging spectroscopy over a broad energy range and at high count-rates. The microcalorimeter technology being baselined is similar to the AXSIO core point source array (PSA), based upon the small-pixel, high count-rate microcalorimeter technology originally developed for solar physics applications[27]. The main difference between the EPE and AXSIO core instrument is that the EPE pixels over-sample the one arc-minute point spread function of the X-ray optic by an even greater amount than the AXSIO PSA, and are thus able to accommodate even greater fluxes, equivalent to over 3 Crab. The TRL of the XMS subsystems ranges from 3 to 5, and ongoing technology development efforts will raise these to TRL 5 to 6 by 2012 and 2013.

The design of the EPE baseline array is a 40x40 array of 3″ microcalorimeter pixels, a 2′ FOV. Each pixel is 145 x 145 μm and is 4.5 μm thick. These are suspended above thermometers that are Transition Edge Sensors (TES). TESs sensors operate by biasing the sensor to a temperature between its superconducting state and resistive state, so that any small change in temperature, such as from the absorption of an X-ray, produces a large change in resistance. In the EPE TES design, the sensor films are deposited directly onto solid silicon substrates so that there is a strong thermal conductance to the heat bath, producing very fast detectors. The figure below shows an image of part of a 32x32 close-packed array of pixels, developed for solar physics, on a pitch of 75 μm.

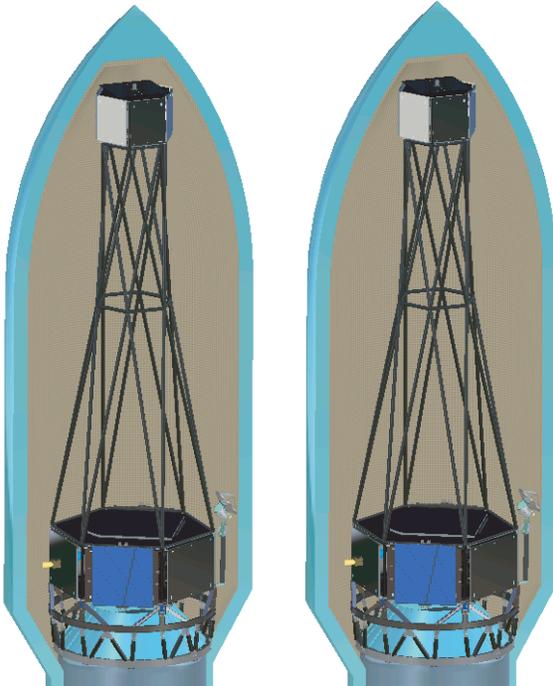

*Figure 14. [Left] Cut-away view of the baseline 10m design in the Falcon 9 fairing. [Right] Same, showing the optional 40m design.*

The best single pixel performance of this type of device is 1.3 eV (FWHM) at 1.5 keV and 1.6 eV (FWHM) at 6 keV, shown in (b) above[28]. These pixels have properties that allow them to accommodate count rates in excess of 300 cps/pixel[27,29]. For EPE we are baselining a pixel size that is two times larger on a side, and a TES size that is scaled by the same factor to produce a similar thermal time-constant. We have conservatively baselined a performance of 2.5 eV (FWHM) for the energy range of 0.2 to 10 keV, and a count rate capability of 150 cps/pixel for the larger pixel size of EPE. When we take into account the shape of the point spread function, the HPD of the X-ray beam covers a 10 pixel radius (or 314 pixels), and the total count rate capability of just over 70k cps for the entire array. These high count-rate detectors require a multiplexed SQUID read-out that uses code division multiplexing[30]. This read-out is more advanced than the time division multiplexing baselined for Athena and previously baselined for IXO.

The EPE XMS is composed of a cryostat that cools an X-ray microcalorimeter array to 50 mK, and the electronics for the detector read-out and for controlling the cooler. Since EPE is being baselined as a class C mission, a cryostat design that has less cryocooler redundancy is considered acceptable (similar to the one originally baselined for IXO, but less complex than the cooler being baselined for AXSIO). In addition, only two models (one flight and one qualification) are necessary, as is the case for ASTRO-H. We have retained the use of redundant cryocooler drive electronics to ensure very high instrument reliability.

## Spacecraft Overview

We have worked with Northrop Grumman to develop concept spacecraft and match them to potential launch vehicles. All subsystems utilize established hardware with substantial flight heritage. Most components are "off-the-shelf." We are designing to NASA 'Class-C' mission specifications and for a 3 year life (5 years expendables) but do include redundancy in systems prone to failure. Both the 10m and 40m concept fit in a Falcon 9 launcher with adequate margin to reach L2. Predicted masses allowing for 22% overall mass growth allowance (30% to 15% on individual subsystems) are 1502 kg and 1809 kg for the 10m and 40m versions respectively. Subsystem level mass rack-ups for both options and details of the spacecraft design are provided in the NGAS appendix.

The L2 orbit facilitates high observational efficiency and provides a stable thermal environment. Articulated solar arrays and a highly reflective cryo-cooler radiator allow nearly complete sky coverage with the exception of within 45 degrees of the sun. EPE carries out observations by pointing at celestial objects for durations of $10^3$–$10^5$ sec. Since the calorimeter is photon counting, longer integrations can be performed by multiple exposures.

EPE consists of four major modules: Instrument, Optical Bench, Spacecraft, and Optics. This architecture facilitates parallel development and integration and test.

**The Instrument Module (IM)**, containing the calorimeter, is shown at the top of the fairing in both figures. With an arcmin PSF there is no need for a focus mechanism. The 10m design is rigid enough that the beam will be remain centered on the detector after launch. The 40m design utilizes stepper motors in the 'tensegrity' structure to steer the beam onto the calorimeter after launch.

**The Optical Bench (OB)** Is near zero CTE graphite and is fixed in the 10m case. The 40m configuration uses a single central Astro Boom which has previously been designed for another mission as is at TRL-7. As the mast deploys, it pulls a pleated shroud that shields the instruments thermally and from stray light.

The **Spacecraft Module (SM)** accommodates the bulk of the spacecraft subsystems including the power; propulsion; RF communications; guidance, navigation, and control; and avionics. This is located within the OM in the 10m case, and within the IM in the 40m case.

The **Optics Module (OM)** includes the foil or MCPO, its sunshade, and the star trackers. The 1.3m diameter foil optic substantially under-fills the SM, but the SM must have a diameter near 3.3m in order to meet the 10 Hz lateral load requirement at the payload adapter fitting. The 4.2m MCPO fits within the fairing with adequate clearance.

## COST ESTIMATING

The **EPE ROM cost is estimated to be $774M (FY11 dollars) including 30% margin and $22.5M for science research grants to the US scientific community.** This cost to NASA covers launch vehicle, mission formulation, development, and operational phases (Phase A-E). The costs for the foil optic and XMS are grass-roots based on actual costs incurred in developing these systems for Suzaku and ASTRO-H. Costs for science, operations and ground data system (GDS) are based on experience with Chandra. The cost for the spacecraft is based on the NGAS massrackup and the JSC SVLCM which is itself derived from NAFCOM. We believe this to be a conservative cost as inputs to this model do not include the Class-C category for EPE. The JSC AMCM model brackets SVLCM for a spacecraft difficulty of low to very low, lending additional credibility to the spacecraft cost. Additional detail is provided below.

**Mirror**: Costs for the foil mirror are scaled at the component level from the actual $2.6M total cost for the ASTRO-H mirror. ASTRO-H housings, gold coating targets, aluminum substrates were $250k, $200k, and $40k; we scale these by 10x to 20x due to the increased area and estimate $2M, $4M, and $0.8M. Forming mandrels were $500k, we estimate $2M acquisition and another $0.5M in pre-manufacture studies due to the larger size; replication mandrel costs acquisition is estimated at $500k. Personnel costs are estimated at $1.25M for assembly technicians (5x1.5years), $0.6M lab engineers (2x2years), $0.9M scientists (2x3years), $0.3M structural/thermal analysis (2x1years), $0.15M design/CAD engineers (1x1year). Total is $13M, we allow for 100% cost growth and budget $26M below. Costs for the MCPO are derived from estimated production costs for the individual glass pieces ($60M) and an assumed equal cost for the design and assembly process. This along with the associated more massive spacecraft adds $169M (including reserves) to the $774M baseline total.

**XMS**: The cost of the complete cooling chain, including cryostat and cryocoolers (qualification and flight model), single string but with redundant cooler drive electronics is $54.8M, items attached to cryostat is $14.3M, detector electronics are $29.3M, the focal plane assembly is $16.8M, support, I&T, s/w is $20.6M for a total of $135.9M.

| Description | WBS # | $M (FY11) | Notes |
|---|---|---|---|
| Management | 1,2,3 | 50 | 15% wrap on optics, detector, and s/c |
| Science | 4 | 59 | Science team starts L-2 years, goes L+4 |
| Optics | 5 | 26 | ASTRO-H scaled and then doubled |
| XMS+cryo | 5 | 136 | Single string cryo, redunant electronics, ICU included in S/C |
| S/C | 6 | 178 | NGAS mass, JSC SVLCM mass only model (~NASCOM FY2011 |
| Ops | 7 | 30 | |
| LV | 8 | 100 | Falcon 9 |
| GDS | 9 | 20 | |
| MSI&T | 10 | 20 | |
| EPO | 11 | 5 | |
| Total, no reserves | | 620 | |
| Reserves | | 149 | 30%, no reserves on GO, LV, EPO |
| Grand Total | | 774 | |
| GO Program | | 22.5 | Assumes 150 $50K grants/year, included in WBS 4 cost above |
| Science program includes GO and $4M/year in science and support by GTO team | | | |

# APPENDIX 1

# APPENDIX 2:

NGAS charts on 10m and 40m configurations, including mass rackups.

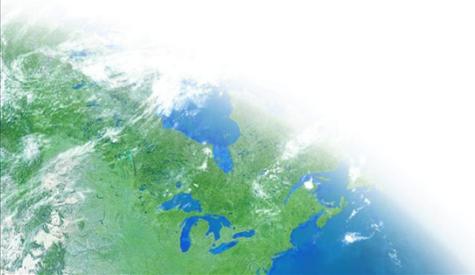

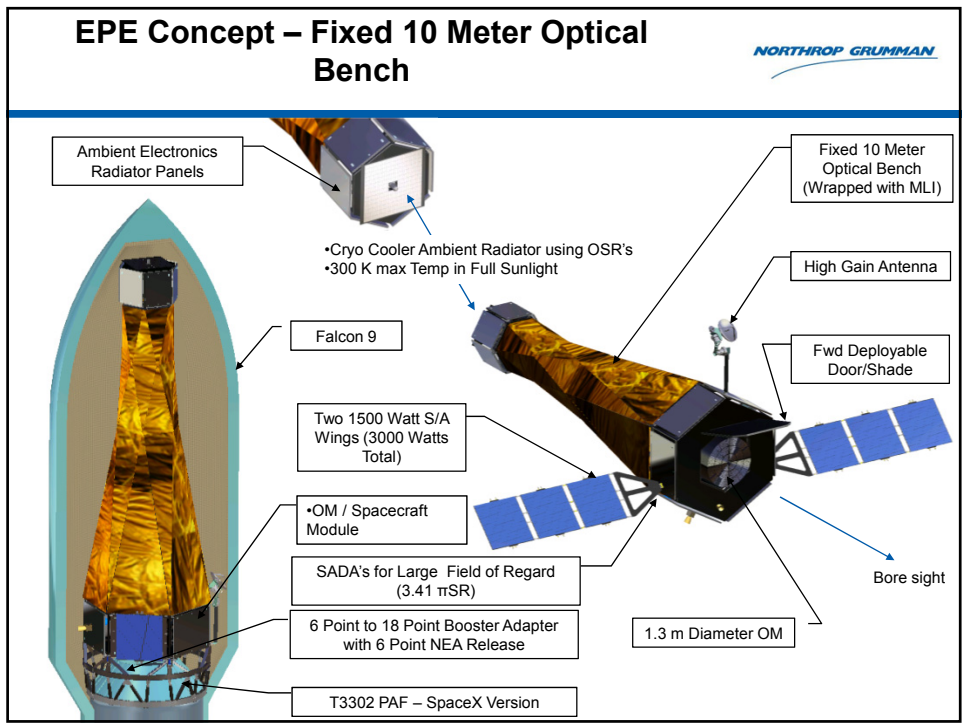

## EPE Concept – Fixed 10 Meter Optical Bench

NORTHROP GRUMMAN

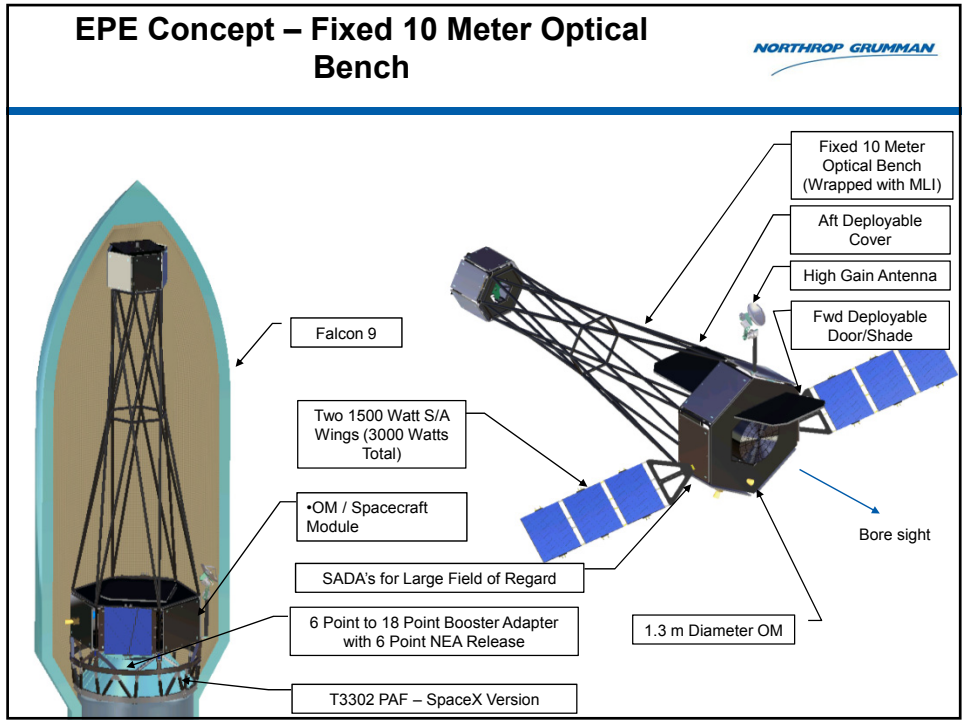

Falcon 9

Two 1500 Watt S/A Wings (3000 Watts Total)

•OM / Spacecraft Module

SADA's for Large Field of Regard

6 Point to 18 Point Booster Adapter with 6 Point NEA Release

T3302 PAF – SpaceX Version

Fixed 10 Meter Optical Bench (Wrapped with MLI)

Aft Deployable Cover

High Gain Antenna

Fwd Deployable Door/Shade

Bore sight

1.3 m Diameter OM

## EPE Fixed 10 Meter Bench – 1.3 Meter Diameter Optical Module Assembly

NORTHROP GRUMMAN

•Scaled Up Astro H Design
 •Focal Length Changed to 10 Meters
 •Additional Mirror Sectors Added

•1.3 m O.D. Stage Added
 •12 Sectors
 •65 Concentric Foil Rings per 30 Degree Sector
 •780 Reflector Foils Segments per Stage

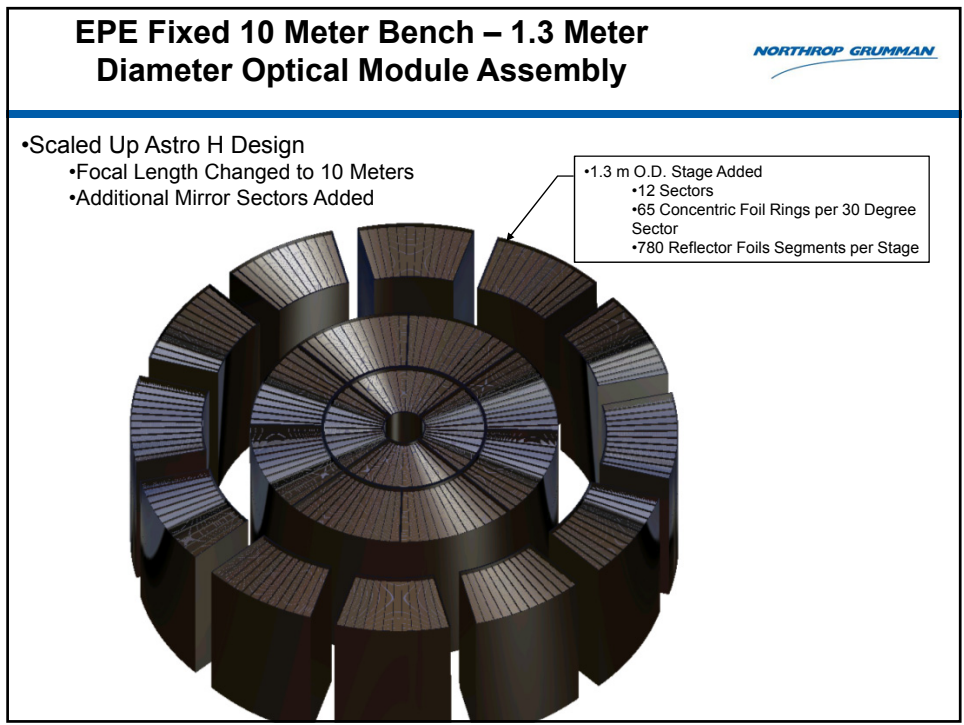

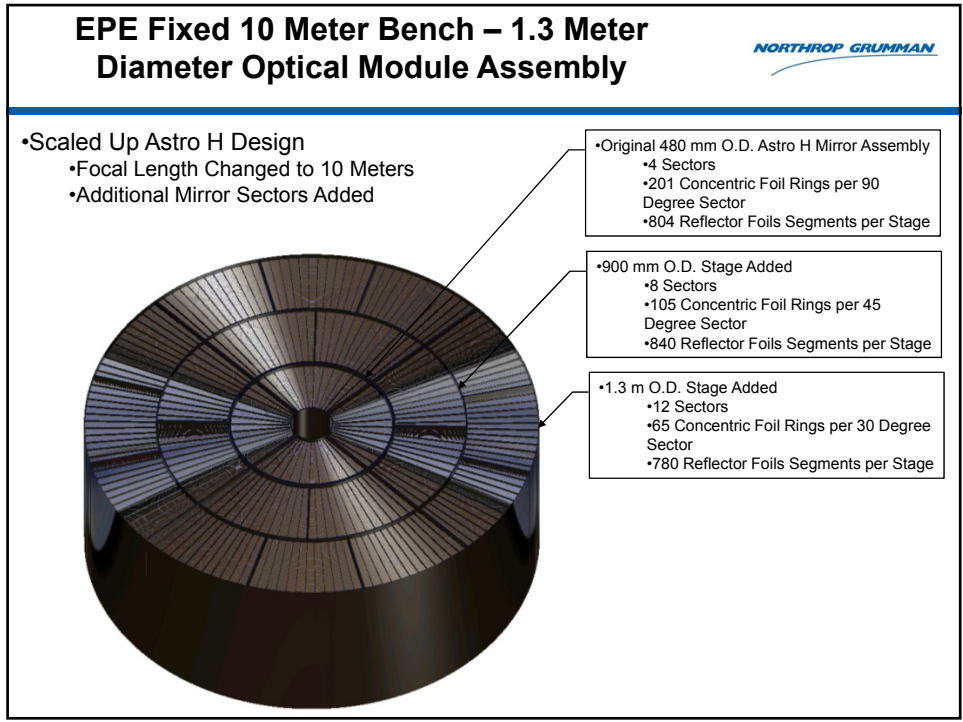

**EPE Fixed 10 Meter Bench – 1.3 Meter Diameter Optical Module Assembly**

NORTHROP GRUMMAN

- Scaled Up Astro H Design
  - Focal Length Changed to 10 Meters
  - Additional Mirror Sectors Added

- Original 480 mm O.D. Astro H Mirror Assembly
  - 4 Sectors
  - 201 Concentric Foil Rings per 90 Degree Sector
  - 804 Reflector Foils Segments per Stage

- 900 mm O.D. Stage Added
  - 8 Sectors
  - 105 Concentric Foil Rings per 45 Degree Sector
  - 840 Reflector Foils Segments per Stage

- 1.3 m O.D. Stage Added
  - 12 Sectors
  - 65 Concentric Foil Rings per 30 Degree Sector
  - 780 Reflector Foils Segments per Stage

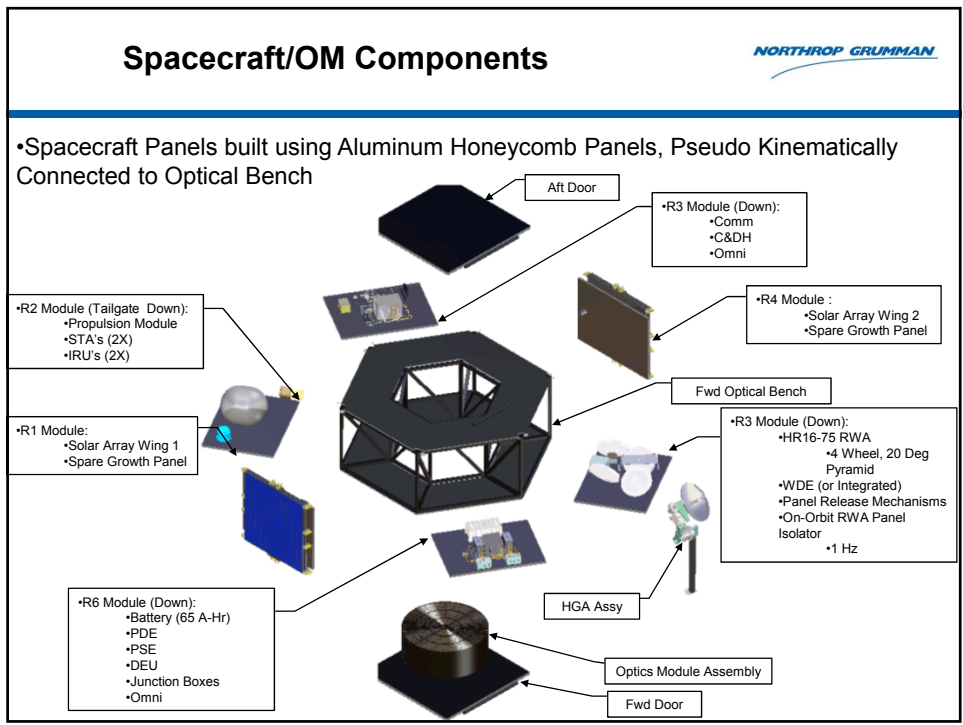

**Spacecraft/OM Components**

NORTHROP GRUMMAN

- Spacecraft Panels built using Aluminum Honeycomb Panels, Pseudo Kinematically Connected to Optical Bench

Aft Door

- R3 Module (Down):
  - Comm
  - C&DH
  - Omni

- R2 Module (Tailgate Down):
  - Propulsion Module
  - STA's (2X)
  - IRU's (2X)

- R4 Module :
  - Solar Array Wing 2
  - Spare Growth Panel

Fwd Optical Bench

- R1 Module:
  - Solar Array Wing 1
  - Spare Growth Panel

- R3 Module (Down):
  - HR16-75 RWA
    - 4 Wheel, 20 Deg Pyramid
  - WDE (or Integrated)
  - Panel Release Mechanisms
  - On-Orbit RWA Panel Isolator
    - 1 Hz

- R6 Module (Down):
  - Battery (65 A-Hr)
  - PDE
  - PSE
  - DEU
  - Junction Boxes
  - Omni

HGA Assy

Optics Module Assembly

Fwd Door

## Instrument Module Equipment

**NORTHROP GRUMMAN**

•Instrument Module Equipment Panels built using Aluminum Honeycomb Panels, Pseudo Kinematically Connected to Optical Bench

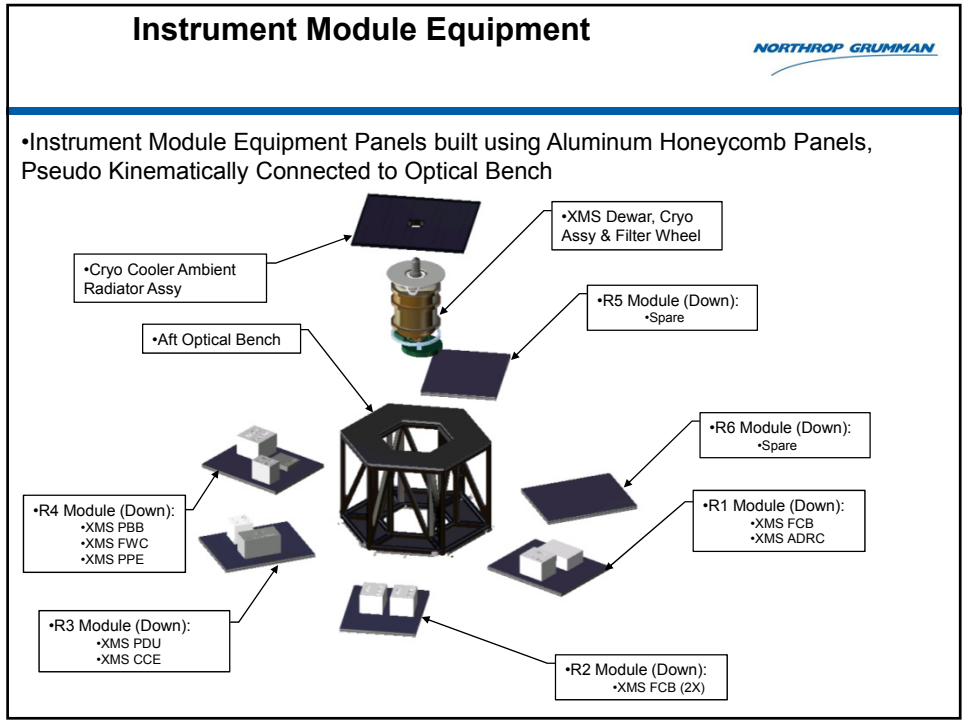

- •Cryo Cooler Ambient Radiator Assy
- •XMS Dewar, Cryo Assy & Filter Wheel
- •R5 Module (Down): •Spare
- •Aft Optical Bench
- •R6 Module (Down): •Spare
- •R4 Module (Down): •XMS PBB •XMS FWC •XMS PPE
- •R1 Module (Down): •XMS FCB •XMS ADRC
- •R3 Module (Down): •XMS PDU •XMS CCE
- •R2 Module (Down): •XMS FCB (2X)

---

## Mass Rollup for Fixed 10 m Design

**NORTHROP GRUMMAN**

| Assembly Level | Unit mass (Kg's) | Qty | Basic Mass (Kg's) | MGA (%) | MGA (Kg's) | Predicted Mass (Kg's) |
|---|---|---|---|---|---|---|
| **EPE Observatory - Wet** | | | 1234.05 | 22% | 268.05 | 1502.10 |
| **Propellant - Monoprop** | | | 78.03 | 21.7% | 16.95 | 94.98 |
| **EPE Payload** | | | 569.68 | 21.7% | 123.83 | 693.51 |
| Optics Module Assy | | | 308.70 | 22.2% | 68.46 | 377.22 |
| FMA | | | 259.84 | 16.6% | 57.38 | 317.21 |
| Bore Sight STA | | | 3.87 | 14.0% | 0.32 | 4.19 |
| Spare Line | 0.00 | 0 | 0.00 | 30.0% | 0.00 | 0.00 |
| Misc Hardware | 2.00 | 1 | 2.00 | 30.0% | 0.60 | 2.60 |
| Harness OM | 3.00 | 1 | 3.00 | 30.0% | 0.90 | 3.90 |
| Deployable Door/Shade | | | 12.55 | 30.0% | 3.77 | 16.32 |
| Deployable Cover - Aft | 20.00 | 1 | 20.00 | 20.0% | 4.00 | 24.00 |
| FMA MLI | 7.50 | 1 | 7.50 | 20.0% | 1.50 | 9.00 |
| Instrument Module Assy | | | 250.27 | 21.3% | 53.42 | 303.68 |
| Instrument Module | | | 243.59 | 21.1% | 51.41 | 295.00 |
| IM Truss | 35.67 | 1 | 35.67 | 30.0% | 10.70 | 46.36 |
| IM Equipment Panels | | | 21.17 | 30.0% | 6.35 | 27.52 |
| XMS Assy | | | 154.75 | 16.3% | 25.24 | 179.99 |
| Payload Accomodation - Misc | 2.00 | 1 | 2.00 | 6.0% | 0.12 | 2.12 |
| Thermal Hardware | 10.00 | 1 | 10.00 | 30.0% | 3.00 | 13.00 |
| IM Harness | 20.00 | 1 | 20.00 | 30.0% | 6.00 | 26.00 |
| Cryo Cooler Ambient Radiator Assy | | | 4.68 | 30.0% | 1.40 | 6.08 |
| Cryo Cooler Ambient Radiator | 2.83 | 1 | 2.83 | 30.0% | 0.85 | 3.68 |
| OSR's | 0.89 | 1 | 0.89 | 30.0% | 0.27 | 1.15 |
| Adhesive | 0.46 | 1 | 0.46 | 30.0% | | |
| Interface Structure | 0.50 | 1 | 0.50 | 30.0% | 0.15 | 0.65 |
| IM MLI | 2.00 | 1 | 2.00 | 30.0% | 0.60 | 2.60 |
| Payload Misc | | | 10.65 | 18.3% | 5.95 | 12.60 |
| **Primary Intergating Structural Assy** | | | 235.51 | 30.0% | 70.65 | 306.17 |
| Forward Hexagonal Box Truss | | | 80.02 | 30.0% | 24.00 | 104.02 |
| Tower Truss | | | 155.50 | 30.0% | 46.65 | 202.14 |
| **EPE Spacecraft (Dry)** | | | 350.83 | 16.1% | 56.61 | 407.45 |
| Spacecraft Secondary Structures & Mechanisms | | | 68.63 | 27.8% | 19.07 | 87.71 |
| Avionics | | | 217.70 | 10.9% | 25.53 | 243.23 |
| Propulsion | | | 26.518 | 7.6% | 2.019 | 28.538 |
| Thermal | | | 33.98 | 26.5% | 8.99 | 42.98 |
| Payload Support Misc | | | 4.00 | 25.0% | 1.00 | 5.00 |

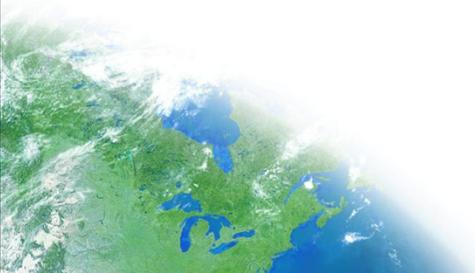

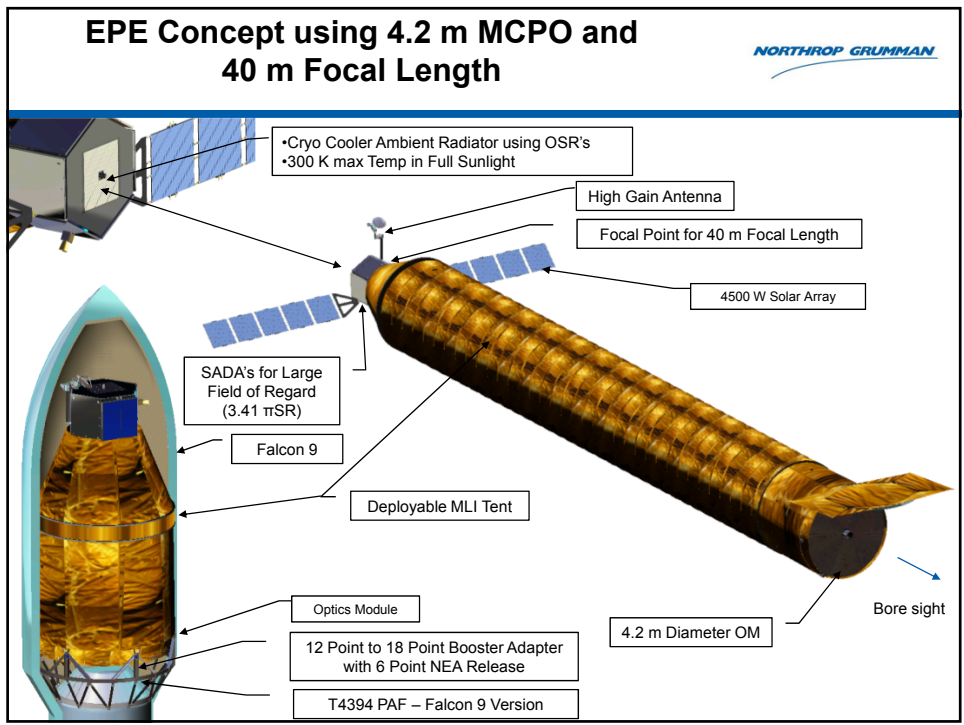

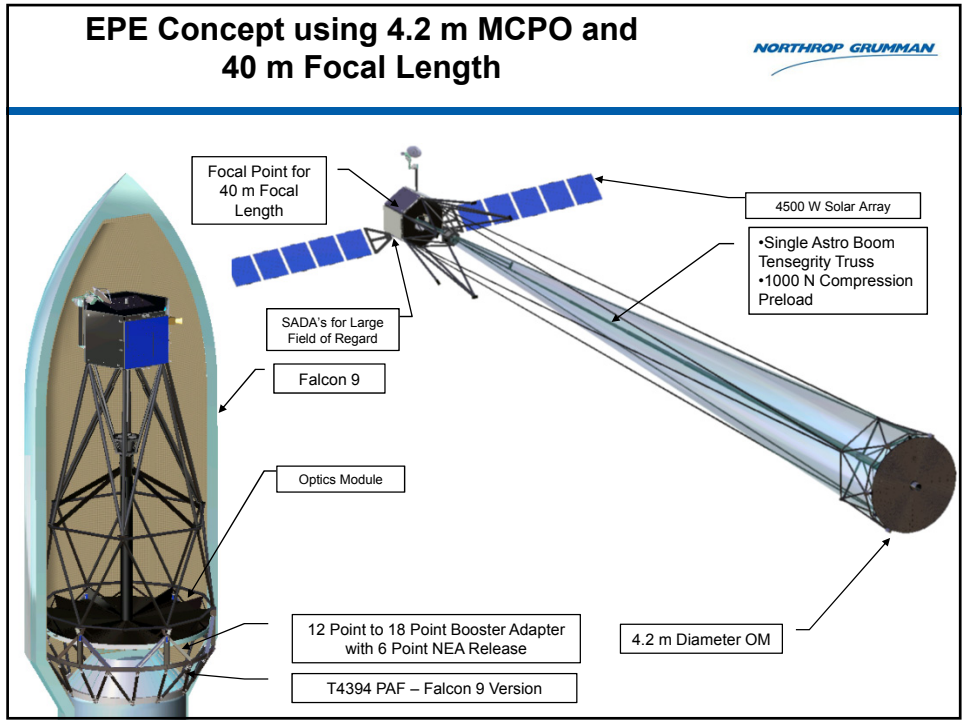

EPE Concept using 4.2 m MCPO and 40 m Focal Length

Focal Point for 40 m Focal Length

4500 W Solar Array

- Single Astro Boom Tensegrity Truss
- 1000 N Compression Preload

SADA's for Large Field of Regard

Falcon 9

Optics Module

12 Point to 18 Point Booster Adapter with 6 Point NEA Release

T4394 PAF – Falcon 9 Version

4.2 m Diameter OM

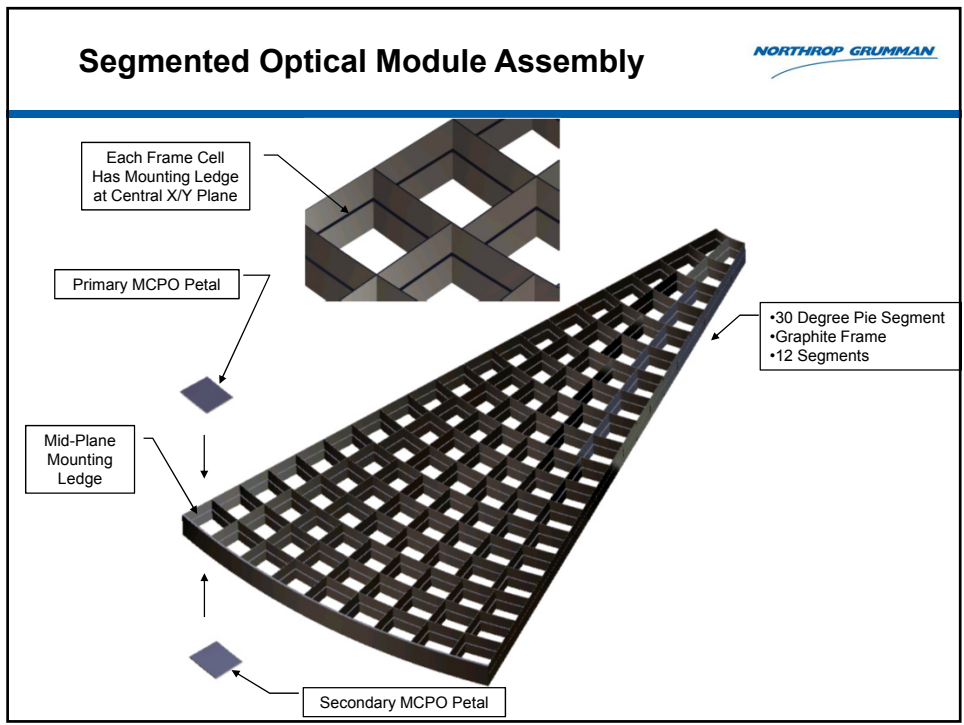

Segmented Optical Module Assembly

Each Frame Cell Has Mounting Ledge at Central X/Y Plane

Primary MCPO Petal

- 30 Degree Pie Segment
- Graphite Frame
- 12 Segments

Mid-Plane Mounting Ledge

Secondary MCPO Petal

## EPE – 4.2 Meter Diameter Optical Module Assembly

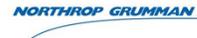

- 1344 Micro channel Plate Optic Petals (Primary)
- 1344 Micro channel Plate Optic Petals (Secondary)
- 19 Petal Rings
- Typical Petal Size = 100 mm
- FMA Structural Fraction ~ 40%

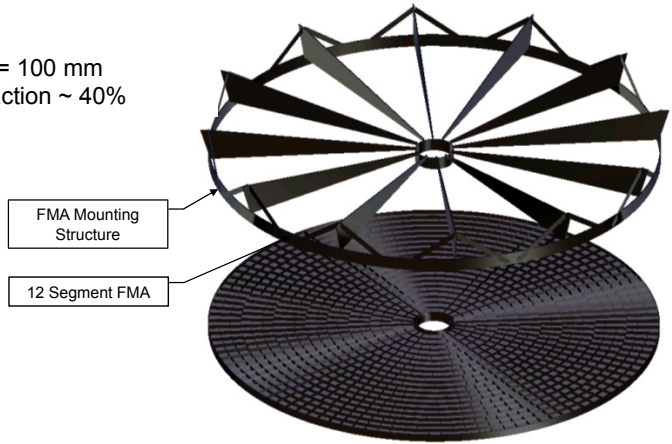

FMA Mounting Structure

12 Segment FMA

## Structural Modes

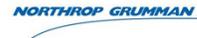

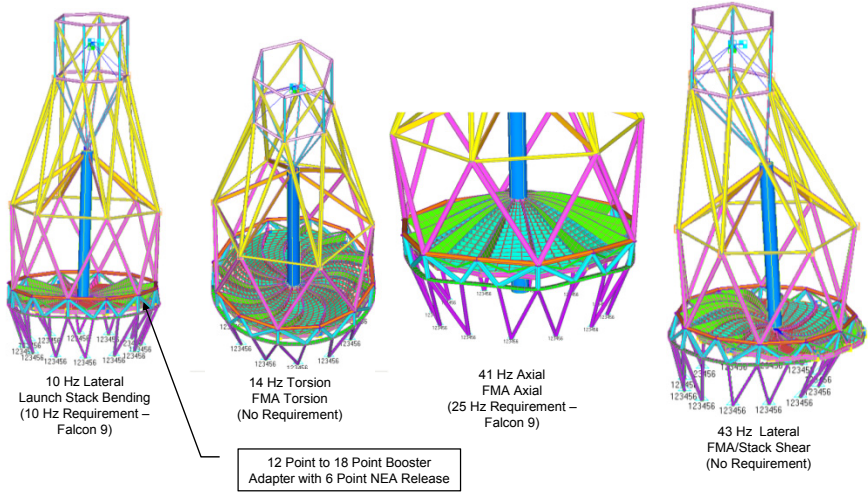

10 Hz Lateral
Launch Stack Bending
(10 Hz Requirement –
Falcon 9)

14 Hz Torsion
FMA Torsion
(No Requirement)

12 Point to 18 Point Booster
Adapter with 6 Point NEA Release

41 Hz Axial
FMA Axial
(25 Hz Requirement –
Falcon 9)

43 Hz Lateral
FMA/Stack Shear
(No Requirement)

# EPE 40 m MCPO Mass Rollup

NORTHROP GRUMMAN

| Assembly Level | | | | | | | | Unit mass (Kg's) | Qty | Basic Mass (Kg's) | MGA (%) | MGA (Kg's) | Predicted Mass (Kg's) |
|---|---|---|---|---|---|---|---|---|---|---|---|---|---|
| Level 1 | Level 2 | Level 3 | Level 4 | Level 5 | Level 6 | Level 7 | Level 8 | | | | | | |
| EPE Observatory - Wet | | | | | | | | | | 1471.82 | 23% | 337.56 | 1809.38 |
| | Propellant - Monoprop | | | | | | | | | 85.16 | 22.3% | 19.53 | 104.70 |
| | EPE Payload | | | | | | | | | 644.46 | 24.0% | 154.54 | 799.00 |
| | | Optics Module Assy | | | | | | | | 330.91 | 25.2% | 83.36 | 414.29 |
| | | | FMA | | | | | | | 275.98 | 18.6% | 70.49 | 346.48 |
| | | | Bore Sight STA | | | | | | | 3.87 | 14.0% | 0.52 | 4.19 |
| | | | Misc Hardwe | | | | | 2.00 | 1 | 2.00 | 30.0% | 0.60 | 2.60 |
| | | | Harness OM | | | | | 3.00 | 1 | 3.00 | 30.0% | 0.90 | 3.80 |
| | | | Deployable DoorShade | | | | | | | 12.55 | 30.0% | 3.77 | 16.32 |
| | | | Deployable Cover - Aft | | | | | 20.00 | 1 | 20.00 | 20.0% | 4.00 | 24.00 |
| | | | FMA MLI | | | | | 7.50 | 1 | 7.50 | 20.0% | 1.50 | 9.00 |
| | | Instrument Module Assy | | | | | | | | 302.90 | 22.8% | 69.21 | 372.11 |
| | | | Instrument Module | | | | | | | 237.87 | 20.9% | 49.70 | 287.57 |
| | | | | IM Truss | | | | 36.45 | 1 | 36.48 | 30.0% | 10.94 | 47.39 |
| | | | | IM Equipment Panels | | | | | | 14.67 | 30.0% | 4.40 | 19.07 |
| | | | | IMIS Assy | | | | | | 154.75 | 16.3% | 25.24 | 179.99 |
| | | | | Payload Accomodation - Misc | | | | 2.00 | 1 | 2.00 | 6.0% | 0.12 | 2.12 |
| | | | | Thermal Hardware | | | | 10.00 | 1 | 10.00 | 30.0% | 3.00 | 13.00 |
| | | | | IM Harness | | | | 20.00 | 1 | 20.00 | 30.0% | 6.00 | 26.00 |
| | | | | Stray Light Baffle Assy | | | | | | 7.50 | 30.0% | 2.25 | 9.75 |
| | | | | IM Truss Adapter Assy | | | | | | 50.85 | 30.0% | 15.26 | 66.11 |
| | | | | Cryo Cooler Ambient Radiator Assy | | | | | | 4.68 | 30.0% | 1.40 | 6.08 |
| | | | | IM MLI | | | | 2.00 | 1 | 2.00 | 30.0% | 0.60 | 2.60 |
| | | | Payload Misc | | | | | | | 10.65 | 18.0% | 1.95 | 12.60 |
| | Deployable Tensegrity Structure Assy | | | | | | | | | 372.61 | 27.8% | 103.74 | 476.35 |
| | | Primary Intergrating Structural Assy - Tensegrity Truss | | | | | | | | 94.91 | 30.0% | 28.47 | 123.38 |
| | | Deployment System - Tensegrity Truss | | | | | | | | 117.31 | 23.1% | 27.10 | 144.45 |
| | | Kapton Blanket Assy - Deployable | | | | | | | | 142.26 | 30.0% | 42.68 | 184.94 |
| | | Deployable Harness Assy | | | | | | | | 18.14 | 30.0% | 5.44 | 23.59 |
| | EPE Spacecraft (Dry) | | | | | | | | | 369.59 | 16.2% | 59.75 | 429.34 |
| | | Spacecraft Secondary Structures & Mechanisms | | | | | | | | 66.35 | 27.7% | 18.36 | 84.61 |
| | | Avionics | | | | | | | | 236.79 | 10.8% | 29.77 | 266.57 |
| | | Propulsion | | | | | | | | 30.518 | 7.6% | 2.318 | 59.530 |
| | | Thermal | | | | | | | | 36.02 | 26.7% | 9.60 | 45.62 |
| | | Payload Support Misc | | | | | | | | 4.00 | 25.0% | 1.00 | 5.00 |